\documentstyle[preprint,aps]{revtex}

\begin{document}
\draft
\title{Hard-core Final State Interaction Effects in Deep Inelastic
 Scattering }
\author {Jaime Besprosvany$^1$ and S. A.  Gurvitz$^{2,3}$}
\address{
$^1$Department of Physics of Complex Systems and
$^2$Department of Particle Physics,\\ Weizmann Institute of
         Science, Rehovot 76100, Israel\\
$^3$TRIUMF, Vancouver, B.C., Canada V6T\ 2A3}
\maketitle
\begin{abstract}
Hard-core final state interaction effects in the
response function are investigated  in the asymptotic limit of
momentum transfer $q\rightarrow\infty$.
The relevant scattering contribution is displayed and
a modification of the otherwise expected free response
is obtained for Bose and Fermi  systems.
A comparison with other treatments of final state interactions is made.
\vskip 5cm
{\it To be published in Phys. Lett. A.}
\end{abstract}

Important information on  Bose and Fermi systems can be
obtained from inclusive scattering of a weak probe in which
energy $\omega$ and momentum $q$ are transferred to these
systems. Such measurements in the large $q$ region
offer the possibility to obtain the single-particle
momentum distribution $n(p)$. Indeed, if for large $q$
the response function $S(q,\omega)$
approaches the impulse approximation (IA), then the momentum distribution
can be readily obtained. In terms of the reduced response
$\phi(q,{\mbox y})\equiv (q/m)S(q,\omega)$
the IA is given as
\begin{eqnarray}
\label {impulse}
\phi^{IA}({\mbox y} )=
\int {d^3p \over (2\pi)^3} n(p)
\delta({\mbox y} -\hat{\mbox{\boldmath $q$}}\cdot{\mbox{\boldmath $p$}})
={1\over 2\pi A}{\mbox {Re}}\int d^3rdx
\exp (i{\mbox y} x)
\rho_1({\mbox{\boldmath $r$}};{\mbox{\boldmath $r$}}+
x\hat{\mbox{\boldmath $q$}}),
\end{eqnarray}
where ${\mbox y} =m\omega/q-q/2$ is a  scaling variable,
$A$  the  number of particles in the system,
$m$ the mass of the particle,
 and
$\rho_1$ is the one-particle density matrix
in coordinate space.  In particular,
an examination of   Eq.  (\ref{impulse})
implies that the IA gives a distinctive delta function peak for
the case  of a momentum distribution containing a condensate
fraction $n_0$, namely  $n(p)=n_{nc}(p)+n_c(p)$,
where $n_c(p)=n_0(2\pi)^3\delta({\mbox{\boldmath $p$}})$.
However, final state interactions (FSI) can modify Eq.
 (\ref{impulse})
and it is of special interest   to study their effect
on the condensate peak.

The treatment of FSI is
a very complicated problem in general and it is desirable to
study the response in the asymptotic limit
$q\rightarrow \infty$, and ${\mbox y} $=const, where the problem is greatly
simplified. Indeed, Eq. (\ref{impulse})
represents the lowest order term in a multiple scattering
series (MSS), i. e.,
$ \phi(q,{\mbox y} )-\phi^{IA}({\mbox y} )=O(v/q)$, where $v$ is
 the potential.
This means that the IA is exact in the asymptotic limit
in the case of regular potentials. In addition,
if the interactions contain a strong repulsive component (hard-core),
only this part has to be  taken into account in the asymptotic limit,
while the regular part will not contribute.

In the following we treat the asymptotic response in terms of a
MSS describing the FSI. Being able to sum over the MSS into  simple
analytical expressions,
 we investigate for both Bose and Fermi systems how
the IA is modified in this limit. In particular,
we explicitly show how the
$\delta$-function condensate peak is smeared due to FSI. We also make
a comparison with other approaches for the response function in
the  asymptotic limit, thus clarifying in this way
the nature of the approximations underlying these.

The exact asymptotic expression for the Green's function entering the
response in the case of hard-core FSI was obtained
in Ref. \cite{Rosen} by use of the  well-known result that the
Schroedinger equation coincides with the equations of
geometrical optics when the energy of the struck particle goes
to infinity. The geometrical optics solution is known
for a wave function
describing hard-core scattering of the knock-on
particle '1' off  spectator particles, '2',...,'$A$'.
It is a plane wave describing particle '1',
modified by a subtraction
of  "shadow" regions behind the scatterers and adding the
reflected waves\cite{Born}. The      resulting
form for the   response $\tilde \phi ({\mbox y})\equiv
{\rm \lim}_{q\rightarrow\infty} \phi (q,{\mbox y})$ is
\begin{eqnarray}
\label {eik}
\tilde\phi({\mbox y})=
{\frac {1}{A!\pi}} {\mbox {Re}}
\int d^2b_1dz_1dz_1^\prime
d^3r_2 ... d^3r_A
\rho_A \theta (z_1^\prime-z_1)
e^{\displaystyle i {\mbox y} (z_1^\prime-z_1)}\prod_{i=2}^{A}
 ( 1-S_{1i}),
\end{eqnarray}
where
\begin{equation}
\label {amp}
S_{1i}=\theta(a-\vert {{\mbox{\boldmath $b$}}_1}-
{{\mbox{\boldmath $b$}}_i} \vert )
  \theta(z_1^\prime-z_i+w_i)
  \theta(w_i-z_1+z_i)
\end{equation}
with $w_i=(a^2-|{{\mbox{\boldmath $b$}}_1}-
{{\mbox{\boldmath $b$}}_i}|^2)^{1/2}$,
${{\mbox{\boldmath $r$}}_1}=({{\mbox{\boldmath $b$}}_1},z_1)$,
${{\mbox{\boldmath $r$}}'_1}=
({{\mbox{\boldmath $b$}}_1},z'_1)$,
and $a$ is the radius of the core.
$\rho_A$ is the $A$-body density matrix defined  as
\begin{eqnarray}
\label {rhoA}
\rho_A\equiv\rho_A({{\mbox{\boldmath $r$}}_1,
{\mbox{\boldmath $r$}}_2,...,
{\mbox{\boldmath $r$}}_A}
;{{\mbox{\boldmath $r$}}'_1,{\mbox{\boldmath $r$}}_2,...,
{\mbox{\boldmath $r$}}_A})=
A! \Psi_0({{\mbox{\boldmath $r$}}_1},{{\mbox{\boldmath $r$}}_2,...,
{\mbox{\boldmath $r$}}_A})
\Psi_0^*({{\mbox{\boldmath $r$}}'_1},{{\mbox{\boldmath $r$}}_2,...,
{\mbox{\boldmath $r$}}_A}).
\end{eqnarray}
where $\Psi_0$ is the ground state wave
function of the $A$-particle system,
normalized in a
box of volume $V$ as
$\langle  \Psi_0 \vert \Psi_0 \rangle=1$.
The corresponding $n$-body density matrices are defined by:
\begin{eqnarray}
\label{den}
\rho_n({{\mbox{\boldmath $r$}}_1,...,{\mbox{\boldmath $r$}}_n;
{\mbox{\boldmath $r$}}'_1,...,
{\mbox{\boldmath $r$}}_n})=
     {1\over(A-n)!}\int d^3r_{n+1}\cdots d^3r_A
\rho_A({{\mbox{\boldmath $r$}}_1,...,{\mbox{\boldmath $r$}}_A;
{{\mbox{\boldmath $r$}}_1}^\prime,...,{\mbox{\boldmath $r$}}_A}).
\end{eqnarray}
Notice that the $\rho_A$ needed is
off-diagonal in the coordinate of the struck
particle ${{\mbox{\boldmath $r$}}_1}$,
and  diagonal in the coordinates of the spectator
particles.

With  its rather simple appearance, Eq. (\ref{eik})
gives the exact account for the
FSI of the knock-on particle with spectators in the asymptotic limit.
It is only the hard-core component of FSI which survives in this limit
 and modifies the
free eikonal Green's function in Eq. (\ref{eik}) by the product
of $(1-S_{1i})$. This product
forbids propagation along the
hard-core shadow configurations (the hatched region in Fig. 1).
By expanding the product in the integrand (\ref{eik})
\begin{equation}
\label {expan}
\prod_{i=2}^A(1-S_{1i})=1-\sum_iS_{1i}+\sum_{i>j}S_{1i}S_{1j}-
\sum_{i>j>k}S_{1i}S_{1j}S_{1k}+\cdots
\end{equation}
one gets parallelly for the response function
\begin{equation}
\label {mss}
\tilde\phi({\mbox y})=\sum_{i=1}^A\tilde\phi_i({\mbox y})
\end{equation}
where the first term, $\tilde\phi_1({\mbox y})$,
corresponds to IA, Eq. (\ref{impulse}) and the higher order
terms represent  a Glauber-type MSS, describing FSI of the
knock-out particle with $A-1$ spectator particles.

As a first instructive example for the application of Eq. (\ref{eik})
we consider the asymptotic response of a system of
$n_0 A$ particles occupying the zero momentum state, as
can be obtained from the Bose condensation phenomenon in an ideal
gas. We take the $A$-body density-matrix in
the  form $\rho_A=\rho_{Ac}^{(1)}=n_0A!/V^A$,
demanding
$ \rho_{1c}({{\mbox{\boldmath $r$}}_1;{\mbox{\boldmath $r$}}'_1})=
 n_0\rho$,
where $\rho =A/V$. By substituting
$\rho_{Ac}^{(1)}$ into Eq. (\ref{eik}), we notice
that its independence
of the particles' coordinates
enables Eq. (\ref{eik}) to be written as
 products of equivalent integrals over the spectator coordinates:
\begin{eqnarray}
\label{eikm}
 \tilde\phi_c^{(1)}({\mbox y})=
{\frac {n_0}{ \pi  A}}
{\mbox {Re}}
 \int dz_1dz_1^\prime d^2 b_1\rho\theta(z_1^\prime-z_1)
\exp \left [ i{\mbox y} (z_1^\prime-z_1) \right ]
{\left [1-\frac{1}{A} \rho \tilde E(z_1^\prime-z_1)\right ] }^{A-1},
\end{eqnarray}
with
\begin{eqnarray}
\label{GTG}
\tilde E(z_1^\prime-z_1)=
\int  d^2b_idz_i\theta(a-\vert {{\mbox{\boldmath $b$}}_1-
{\mbox{\boldmath $b$}}_i} \vert )
  \theta(z_1^\prime-z_i+w_i)
  \theta(w_i-z_1+z_i)
                       \nonumber \\
=2\pi\int_0^a   db_{1i} b_{1i}
(z_1^\prime-z_1+2\sqrt{a^2-b_{1i}^2})
=\pi a^2(z_1^\prime-z_1+{4a\over 3}),
\end{eqnarray}
and $ b_{1i}=
|{{\mbox{\boldmath $b$}}_1}-
{{\mbox{\boldmath $b$}}_i}|$.
The function $\tilde E$
 depends only on   $ x=z_1^\prime-z_1  $
and then Eq. (\ref{eikm}) may be rewritten as
\begin{eqnarray}
\label{eikcof}
 \tilde\phi_c^{(1)}({\mbox y})=
{\frac {n_0}{ \pi  }}
{\mbox {Re}}
\int_0^\infty dx
e^{\displaystyle  i{\mbox y} x }
\left [1- \frac{1}{A} \rho \pi a^2( x+{4a\over 3})\right ]^{A-1},
\end{eqnarray}
By expanding the integrand in Eq. (\ref{eikcof}) using the binomial
 formula
one explicitly reproduces the IA term and the MSS, Eq. (\ref{mss}).
In the limit $A\rightarrow\infty$   Eq. (\ref{eikcof})
goes over to
\begin{eqnarray}
\label{eikfin}
 \tilde\phi_c^{(1)}({\mbox y})= \frac {n_0}{ \pi  }
   e^{{-4\pi \rho a^3\over 3}}
    {\mbox {Re}}  \int^{\infty}_0 dx
       \exp \left [ i{\mbox y} x- \rho \sigma x
\right ] =\frac {n_0}{ \pi  }
  e^{{-4\pi \rho a^3\over 3 }}{\sigma\rho\over
   (\sigma\rho)^2+{\mbox y}^2},
\end{eqnarray}
where $\sigma=\pi a^2$.

This result explicitly shows smearing of the $\delta$-function
 peak due to
FSI, first suggested  by Hohenberg and Platzman\cite{PlatH}.
Notice that the $\delta$-function peak survives in the IA
 for finite $A$, and disappears only if
$A\rightarrow\infty$. This explains why  smearing of the
 condensate did not
occur in a truncated MSS\cite{rinat}. Our treatment implies
that it is  not possible  to determine
 a finite number of
collisions leading  to the extinction of the singularity as
imprecisely
stated in Ref. \cite{RinCar}.\footnote{Since the main
contribution of the integral in    Eq. (\ref{eikfin})
 comes from any interval
$\left [0,\bar x \right ]$ with $\bar x >>1/( \rho \sigma)$,
 the response could be approximated by contributions from
a finite number of collision terms
$N\sim \rho \sigma \bar x$ (replacing $e^{-\rho \sigma x}$ in the
integrand in Eq.  (\ref{eikfin})
 by its  Taylor expansion to  $N$th order).
 However, contributions from the
interval $\left [\bar x, \infty  \right]$ remain singular
unless one takes $N\rightarrow\infty$  which corresponds to considering
 an  {\it infinite}  number of collisions.}
 In addition, unlike Ref. \cite{RinCar},
in our treatment
the role of  the hard-core in the smearing
is specific to the asymptotic limit. However, at finite $q$
regular interactions can also produce the effect\cite{BES}.

Eqs. (\ref{eikcof}), (\ref{eikfin})
give the asymptotic response for systems with a constant
$\rho_A$  in space.
Actually,   such $\rho_A$ cannot be produced
because of the presence of configurations with  particles' cores
interceptions (square-hatched region in Fig. 1).
In fact, the resulting Lorentzian
 response in Eq. (\ref{eikfin}) violates
sum rules, and
in particular, $\int\tilde\phi_c^{(1)}({\mbox y})d{\mbox y}\neq n_0 $,
and in fact, the latter integral
and $\int\tilde\phi_c^{(1)}({\mbox y}){\mbox y}^2d{\mbox y}$ diverge.
These problems are overcome by modifying the density-matrix
$\rho_{Ac}^{(1)}$,
imposing the hard-core constraint. Specifically,  we consider
the $A$-body density-matrix
\begin{eqnarray}
\label{facrho}
\rho_{Ac}^{(2)}
=\rho_{Ac}^{(1)}\eta C({\mbox{\boldmath $r$}}_i,
{\mbox{\boldmath $r$}}_i^\prime),
\end{eqnarray}
where
$C({\mbox{\boldmath $r$}}_i,{\mbox{\boldmath $r$}}_i^\prime)=
\prod_{i=2}^A\theta(r_{i1}-a)\theta(r_{i1}^\prime -a)$,
and $r_{1i}=|{{\mbox{\boldmath $r$}}_i-{\mbox{\boldmath $r$}}_1}|$,
${r'}_{1i}=|{{\mbox{\boldmath $r$}}_i-{\mbox{\boldmath $r$}}_1^\prime}|$.
Here the normalization factor $\eta$ is included,
in order to to satisfy the condition for the one-body density
$ \rho_{1c}= n_0\rho$, Eq. (\ref{den}). We obtain
 \begin{eqnarray}
 \label{wavenor}
 \eta(x)=\left (1-u(x)/V\right )^{1-A},
\end{eqnarray}
where
 \begin{eqnarray}
 \label{uavenor}
 u(x)=
 \frac {8}{3}\pi a^3+
( \pi a^2x-\frac{1}{12}\pi x^3
 -\frac {4}{3}\pi a^3)\theta(2a-x).
\end{eqnarray}
Configurations with particles' core interception
for the hit and spectator particles, (Fig. 1),
are actually excluded automatically by the
$(1-S_{1i})$  factors in Eq. (\ref{eik}).
Explicitly, when the latter Eq. is evaluated by substituting
into it $\rho_{Ac}^{(2)}$ given
in Eq. (\ref{facrho}),  the factor
$C({\mbox{\boldmath $r$}}_i,{\mbox{\boldmath $r$}}_i^\prime)$
is redundant.  For this density-matrix
it immediately follows
 \begin{eqnarray}
 \label{eqres}
\tilde\phi_c^{(2)}({\mbox y})=
\frac {n_0}{\pi}
  e^{{-4\pi \rho a^3\over 3}}{\mbox {Re}}
 \int^{\infty}_0 dx    \exp \left [ i{\mbox y} x- \rho (\sigma x
-u(x) )
\right ]
\end{eqnarray}
The behavior of
$\tilde\phi_c^{(2)}({\mbox y})$, being
 other than Lorentzian at large
$|{\mbox y}|$, allows for the (partly asymptotic)
 sum rules to be satisfied. Indeed
one obtains from Eq. (\ref{eqres})
 \begin{eqnarray}
 \label{rules}
 \int d{\mbox y}\tilde \phi^{(2)}({\mbox y})  & = &n_0   \\
 \int d{\mbox y}\tilde \phi^{(2)}({\mbox y}) {\mbox y}&=& 0    \\
 \int d{\mbox y}\tilde \phi^{(2)}({\mbox y}){\mbox y}^2&=&
 \frac{ 2  m}{3} <K>
\end{eqnarray}
where
\begin{equation}
\label{kinen}
<K>=-\left.\frac{1}{2m}\frac{d^2}{dx^2}\rho_1^{(2)}(0,x)\right |_
{x\rightarrow 0}
\end{equation}
is the average kinetic energy.

At low densities or at low momenta
  $|{\mbox y} a|\leq 1$
the second term in Eq. (\ref{uavenor}) can    be neglected.
In  the limit $A\rightarrow\infty$,
$\eta\simeq e^{{8\pi \rho a^3\over 3 }}$, and this Eq. gives
\begin{eqnarray}
\label{eikfinm}
 \tilde\phi_c^{(2)}({\mbox y})\simeq
\frac {n_0}{ \pi  }
  e^{{4\pi \rho a^3\over 3 }}{\sigma\rho\over {\displaystyle
(\sigma\rho)^2+{\mbox y}^2}},
\end{eqnarray}
which differs from the previous result in Eq. (\ref{eikfin}) by the
sign in the  exponent. We
notice also the higher peak modification as compared
with  the Lorentzian
form that can be  obtained from Ref. \cite{Plat} for a structureless
fluid.

The above simple model for the response
 reproduces asymptotic sum rules and
embodies the smearing effect of the hard-core interaction by an
infinite number of spectators. Other
  models with more realistic density matrices,
 are now shown to share yet  the same basic
underlying structure and behavior.

The wave function in the product form $\Psi_0=\prod_{i\not =j}f(r_{ij})
/V^{A/2}$  describes more suitably
  the correlations between the particles and is
often used in calculations, (see for instance\cite{car}).
The function $f(r_{ij})$ is directly related to the pair-correlation
function and tends to $1$ for $r_{ij}$ larger than the correlation
length. Correspondingly, the off-diagonal density matrix can be
written as
\begin{equation}
\label{newd}
\rho_A^{(pr)}=\frac{A!}{V^A}\prod_if(r_{1i})f({r'}_{1i})
\prod_{j,k\not =1}f^2(r_{jk})
\end{equation}
Now we substitute Eq. (\ref{newd}) into Eq. (\ref{eik}) and use the
expansion (\ref{expan}), (\ref{mss}). The first term, $\tilde\phi_1$
is the IA, Eq. (\ref{impulse}). For the second term one finds
\begin{equation}
\label{sec}
\tilde\phi_2^{(pr)}({\mbox y} )=
-{1\over \pi A}{\mbox {Re}}\int_0^{\infty}dx\int d^3r_1
\exp (i{\mbox y}x)
E({{\mbox{\boldmath $r$}}_1},
{{\mbox{\boldmath $r$}}_1}+x\hat{\mbox{\boldmath $z$}})
\end{equation}
where
$\hat{\mbox{\boldmath $z$}}\equiv{\mbox{\boldmath $q$}}/q$, $x=z'_1-z_1$, and
\begin{equation}
\label {assres}
E({{\mbox{\boldmath $r$}}_1,{\mbox{\boldmath $r$}}'_1})=\int   d^3r_2
\rho_2^{(pr)}({{\mbox{\boldmath $r$}}_1,
{\mbox{\boldmath $r$}}_2};{{\mbox{\boldmath $r$}}'_1,
{\mbox{\boldmath $r$}}_2})
\theta(a-\vert {{\mbox{\boldmath $b$}}_1-
{\mbox{\boldmath $b$}}_2} \vert )
  \theta(z_1^\prime-z_2+w_2) \theta(w_2-z_1+z_2).
\end{equation}
Here $\rho_2$ is two-body density matrix, Eq. (\ref{den}).
Notice that the $\theta$-functions in Eq. (\ref{assres})
reduce the $r_2$-integration to the hatched region shown in Fig. 1.

For the third order term, $\tilde\phi_3$  also the integration
over the  spectator coordinates
${{\mbox{\boldmath $r$}}_2}$, ${{\mbox{\boldmath $r$}}_3}$,
 is restricted by the same
hatched region (Fig. 1). While the final expression  now includes
the three-body density matrix
it can yet be approximately reduced to one involving
two-body density matrices by noticing  that the function  $f(r_{23})$,
describing two-body correlations in Eq. (\ref{newd}), is close to one
except for the region of small $r_{23}$.
For $x$ large enough  (larger than the correlation length),
$r_{23}$ will also be large
for most values of the spectator coordinates,
${{\mbox{\boldmath $r$}}_2}$, ${{\mbox{\boldmath $r$}}_3}$,
in their non-vanishing integration region (Fig. 1),
and  one can replace $f(r_{23})$
by 1 in the corresponding integral for $\tilde\phi_3$.
 One  obtains from Eqs. (\ref{den}), (\ref{newd})
\begin{equation}
\label{rel}
\int d^3r_4\cdots d^3r_A
\rho_A^{(pr)}({{\mbox{\boldmath $r$}}_1,...,{\mbox{\boldmath $r$}}_A;
{{\mbox{\boldmath $r$}}_1}^\prime,...,{\mbox{\boldmath $r$}}_A})=
\frac{\rho_2^{(pr)}({{\mbox{\boldmath $r$}}_1,
{\mbox{\boldmath $r$}}_2};{{\mbox{\boldmath $r$}}'_1,
{\mbox{\boldmath $r$}}_2})
\rho_2^{(pr)}({{\mbox{\boldmath $r$}}_1,
{\mbox{\boldmath $r$}}_3};{{\mbox{\boldmath $r$}}'_1,
{\mbox{\boldmath $r$}}_3})}{(A-1)\rho_1^{(pr)}
({{\mbox{\boldmath $r$}}_1};
{{\mbox{\boldmath $r$}}'_1})}
\end{equation}
Using this result one gets for $\tilde\phi_3$
\begin{equation}
\label{third}
\tilde\phi_3^{(pr)}({\mbox y} )\simeq
{1\over \pi A}{\mbox {Re}}\int_0^{\infty}dx\int d^3r_1
\exp (i{\mbox y}x)
\frac{(A-2)E^2({{\mbox{\boldmath $r$}}_1},
{{\mbox{\boldmath $r$}}_1}+x\hat{\mbox{\boldmath $z$}})}{2
\rho_1^{(pr)}({{\mbox{\boldmath $r$}}_1},
{{\mbox{\boldmath $r$}}_1}+x\hat{\mbox{\boldmath $z$}})}
\end{equation}

Applying the same approximation for higher order terms we finally
obtain for the asymptotic response
\begin{eqnarray}
\label {asres}
\tilde\phi^{(pr)}({\mbox y} )=
{1\over \pi A}{\mbox {Re}}\int_0^{\infty}dx\int d^3r_1
\rho_1^{(pr)}({\mbox{\boldmath $r$}}_1;{\mbox{\boldmath $r$}}_1+
x\hat{\mbox{\boldmath $z$}})
\exp (i{\mbox y}x)\left [ 1-{E({{\mbox{\boldmath $r$}}_1},
{{\mbox{\boldmath $r$}}_1}+x\hat{\mbox{\boldmath $z$}})
\over (A-1)\rho_1^{(pr)}({{\mbox{\boldmath $r$}}_1};
{{\mbox{\boldmath $r$}}_1}+x\hat{\mbox{\boldmath $z$}})}
\right ]^{A-1}
\end{eqnarray}

Here too, similarly to  Eq. (\ref{eikm}),
the FSI modifies the IA result,
Eq. (\ref{impulse}), by an additional correction
factor in the integrand. If $\rho_1
({\mbox{\boldmath $r$}}_1;{\mbox{\boldmath $r$}}_1+
x\hat{\mbox{\boldmath $z$}})\rightarrow {\mbox {const}}$ for
$x\rightarrow\infty$, the FSI correction factor would smear the
$\delta$-function peak in the response in the limit of $A\rightarrow
\infty$, just in the same way as in the  previous example. Notice that
the integrand (\ref{asres}) is exact for $x\rightarrow\infty$, and
therefore the smearing would always occur in the asymptotic limit.

Other treatments of the response can be compared with
our asymptotic results.
In 1973 Gersch and Rodriguez (GR)\cite{Geri}
obtained the following formula for the response,
 using the first cumulant term:
\begin{eqnarray}
\label{GERI}
\phi^{(GR)}({\mbox y} )=
\frac{1}{\pi\rho }{\mbox {Re}} \int_0^\infty dx\rho_1(0;x)
\exp\left [i{\mbox y} x+
\int d^3r\frac{\rho_2({{\mbox{\boldmath $r$}}},0;
{{\mbox{\boldmath $r$}}}+
x\hat{\mbox{\boldmath $z$}}, 0)}{\rho_1(0;x)}
 \Omega^{(GR)}({{\mbox{\boldmath $r$}}},
{{\mbox{\boldmath $r$}}}+x\hat{\mbox{\boldmath $z$}})\right ],
\end{eqnarray}
where
\begin{eqnarray}
\label{Gerigr}
 \Omega^{(GR)}({{\mbox{\boldmath $r$}},{\mbox{\boldmath $r$}}'}+
x\hat{\mbox{\boldmath $z$}})=\exp\left [
 -\frac{im}{q}\int_0^x dx' [ v({{\mbox{\boldmath $r$}}}+
x'\hat{\mbox{\boldmath $z$}})-
 v({{\mbox{\boldmath $r$}}}+x\hat{\mbox{\boldmath $z$}}) ]\right ]-1,
\end{eqnarray}
and
$v$ is the inter-particle potential. Here translational invariance
is assumed, so that
$\rho_1({{\mbox{\boldmath $r$}}_1;{\mbox{\boldmath $r$}}'_1})=
\rho_1(0;x)$, and $\rho_2
({{\mbox{\boldmath $r$}}_1,{\mbox{\boldmath $r$}}_2;
{\mbox{\boldmath $r$}}'_1,{\mbox{\boldmath $r$}}_2})
=\rho_2({{\mbox{\boldmath $r$}}},0;{{\mbox{\boldmath $r$}}}+
x\hat{\mbox{\boldmath $z$}},0)$, where
${\mbox{\boldmath $r$}}={\mbox{\boldmath $r$}}_1-
{\mbox{\boldmath $r$}}_2$, and
${\mbox{\boldmath $r$}}'={\mbox{\boldmath $r$}}'_1-
{\mbox{\boldmath $r$}}_2$.

Although the original GR treatment was not intended for singular
interactions\cite{Geri}, Eq. (\ref{GERI}) can be formally applied
in the case of a hard-core interaction by taking
$v\rightarrow \infty$ when the particles are within their
hard-core range. We
consider Eq. (\ref{Gerigr}) in the limit  $q\rightarrow\infty$
and we take into account only the hard-core component of $v$ in Eq.
(\ref{Gerigr}) (since the regular component does not contribute
in this limit).
Then, neglecting the second term in Eq. (\ref{Gerigr}) (since
$\rho_2=0$ for $|{\mbox{\boldmath $r$}}+x\hat{\mbox{\boldmath $z$}}|<a$),
one obtains
\begin{eqnarray}
\label{GER}
\Omega^{(GR)}({\mbox{\boldmath $r$}},{\mbox{\boldmath $r$}}+
x\hat{\mbox{\boldmath $z$}})=
-\theta (a- b)\theta (z+x+w)\theta (w-z).
\end{eqnarray}
Substituting Eq. (\ref{GER}) into Eq. (\ref{GERI}) we find
that in the limit $A\rightarrow\infty$
this equation coincides with our result,
 Eqs. (\ref{assres}), (\ref{asres})
 (assuming
translational invariance for the density matrices $\rho_1$, $\rho_2$).
We thus obtain that the GR formula
reproduces the asymptotic limit, Eq. (\ref{eik}).
Since our result for the FSI correction factor is the exact one
only for $x\rightarrow\infty$, the same is  true for the GR formula.
Also the use of the product form for the density matrix $\rho_A$,
which takes into account only the correlation between the struck
and spectator particles and neglects others\cite{Koh},
would reproduce Eq. (\ref{asres}) and therefore the GR formula
 (\ref{GERI}).
In this connection it is interesting to note
that the same product form\cite{Koh} for $\rho_A$,
is essential in an alternative
derivation of the GR result for regular interactions
using a modified eikonal formula\cite{BES},\cite{Jaime}.

Another treatment of FSI in the response
function, hard-core perturbation theory (HCPT)
has been proposed by Silver\cite{Silver}.
One readily shows
that in the asymptotic limit HCPT and GR lead to different
results\cite{Silver}.
Indeed,  HPCT  response  can be written in the form
of Eq. (\ref{GERI}), where
\begin{eqnarray}
\label{SIL}
\Omega^{(HCPT)}=-\theta (a-b)\theta (z+x)
\end{eqnarray}
(which was used together with $\rho_2/\rho^2=
\theta(r-a)$ in the comparison
with GR),
 is different
from $\Omega^{(GR)}$ as given by Eq. (\ref{GER}).
It thus follows from the above consideration
that HCPT does not reproduce the asymptotic limit for
 $x\rightarrow\infty$.

Next, we discuss the application of our results to the  response
of Fermi systems  interacting through
a hard-core.  Weinstein and Negele\cite{WEI}
studied numerically  such  systems.
 They considered
the asymptotic limit for the response function and found
considerable deviations from the IA at $|{\mbox y}|>k_F$, where $k_F$ is
the Fermi momentum.
We now derive from  our approach a simple
analytical expression for the asymptotic response of a Fermi gas
with hard-core interaction, at small $|{\mbox y}|$.

Consider the limit of
$4\pi\rho a^3/3\ll 1$ (the hard-core volume is small
with respect to the volume per particle). In this case
the Fermi gas momentum distribution, $n_F(p)=\frac{6 \pi^2}{k_F^3}
\theta  (k_F-p)$,
is not essentially modified by  hard-core interactions.
However, the hard-core
inevitably generates high momentum components in $n(p)$.
Therefore the
application of our approach
will  be reliable in the region $|{\mbox y}|<k_F$.
The corresponding one-body density matrix is
\begin{eqnarray}
\label{denf}
\rho_{1F}(0;x)=3\rho\frac{j_1(k_Fx)}{k_Fx}.
\end{eqnarray}
We can consequently write the Fermi gas $A$-particle density matrix as
\begin{eqnarray}
\label{denfa}
 \rho_{AF}=\frac{A!}{\rho V^A}\rho_{1F}(x).
\end{eqnarray}
Substituting Eq. (\ref{denfa}) into Eq. (\ref{eik}) and taking the limit
$A\rightarrow\infty$ we find
\begin{eqnarray}
\label{fermi}
\tilde
\phi_F({\mbox y} )&=&\frac{3}{\pi}{\mbox {Re}}
\int_0^{\infty}\frac{j_1(k_Fx)}{k_Fx}
\exp (i{\mbox y} x-\rho\sigma x)dx\nonumber \\
&=&\frac{3}{4\pi k_F^3}{\mbox {Im}}
\left\{[ ({\mbox y} +i\sigma\rho )^2-k_F^2]
\ln\frac{{\mbox y} +i\sigma\rho +k_F}{{\mbox y}
 +i\sigma\rho -k_F}\right\}-
\frac{3\sigma\rho}{2\pi k_F^2}.
\end{eqnarray}
Notice that in the limit
$\sigma =\pi a^2\rightarrow 0$, Eq. (\ref{fermi})
gives the well known IA result for the Fermi gas response,
$\phi_F^{IA}({\mbox y} )=3(k_F^2-{\mbox y}^2)
\theta (k_F-|{\mbox y}| )/4k_F^3$.

In Fig. 2 we plot the resulting Fermi gas response,
 Eq. (\ref{fermi}), for
different values of the hard-core radius.
We take nuclear matter as an example      with $k_F=1.33$ fm$^{-1}$ and
$\rho =2k_F^3/3\pi^2$. The solid line shows the IA result ($a=0$), and
the dot-dashed and dashed lines correspond to $a$=0.5 fm and $a$=0.8 fm,
respectively. The ratio of the hard-core volume to $1/\rho$
is 0.084 for $a$=0.5 fm, and 0.34 for $a$=0.8 fm. Therefore
the use of the Fermi momentum distribution
in Eq. \ref{denfa} is still approximately
correct. We find that the response considerably deviates from
the IA.
One also notices
the disappearance of the singularity in the  response
at $|{\mbox y}| =k_F$ due to hard-core FSI.

It is also interesting to note the sizable effect from the hard-core
FSI in the region of the quasi-elastic peak. For $a$=0.5 fm the response
is reduced by about 10{\%}. Although such a reduction has been the result
of a pure hard-core FSI, a similar effect of FSI would be also
expected for a
strong repulsive interaction, provided the potential is much
larger than the energy of the recoil particle for distances smaller
than $a$.

In summary, we have investigated the effect of  hard-core interactions
on the response function obtained from inclusive scattering,
around the  quasielastic peak.
 Relevant models of the density matrix describing
systems of particles interacting through a hard-core were constructed.
By using them the response
was shown to be modified and smeared as compared with the otherwise
expected
IA form in the asymptotic limit, both for bosons and
fermions. New analytical formulae were obtained for
the response of these systems  reproducing quite generally
the asymptotic behaviour of the response and that
 satisfy the asymptotic sum rules.

{\bf Acknowledgements}\par

We thank A. S. Rinat for discussions and suggestions in the writing.

\clearpage
{\bf Figure Captions}. \par
Fig 1. Hard-core shadow configuration in space
 (${\mbox{\boldmath $r$}}_i$)
for the spectator particle, produced by the Green's function in
Eq. (\ref{eik}) (hatched region).
Square-hatched circles show the regions of the hard-cores' interception.

Fig. 2. Asymptotic response for a Fermi-gas in the
impulse approximation  (solid), and with  hard-core
 final state interactions
 corrections with core radius $a$=0.5 fm (dot-dashes), and
 $a$=0.8 fm (dashes).

\clearpage

\begin{thebibliography}{99}
\bibitem{Rosen}{S. A. Gurvitz, A. S. Rinat, and  R. Rosenfelder,
Phys. Rev. C  {\bf 40}  (1989) 1363.}
\bibitem{Born}{M. Born and E. Wolf, {\it Principles of Optics} (Pergamon,
Oxford, 1970), Chap. 3.}
\bibitem{PlatH}{P. C. Hohenberg and P. M. Platzman,  Phys. Rev.
{\bf 152}  (1966) 198.}
\bibitem{rinat}{A. S. Rinat and M. F. Taragin, Phys. Rev. B {\bf 41}
(1990) 4247.}
\bibitem{RinCar}{C. Carraro, A. S. Rinat,
 Phys. Rev. B  {\bf 45}  (1992) 2945.}
\bibitem{BES}{J. Besprosvany,  Phys. Rev. B {\bf 46}   (1992) 14226.
}
\bibitem{Plat}{P. M. Platzman, N. Tzoar, Phys. Rev. B {\bf 30}
 (1984) 6397.}
\bibitem{car}{C. Carraro and S.E. Koonin,  Phys. Rev. Lett.
{\bf 65}  (1990) 2792.}
\bibitem{Koh}{A. Kohama, K. Yazaki and R. Seki, Nucl. Phys.  A {\bf 536},
(1992) 716.}
\bibitem{Geri}{ H. A. Gersch and   L. J.  Rodriguez,
Phys. Rev.  A {\bf 8}   (1973) 905.}
\bibitem{Jaime}{J. Besprosvany, Phys. Rev. B {\bf 43}  (1991) 10070. }
\bibitem{Silver}{R. N. Silver,  Phys. Rev. B  {\bf 38}  (1988) 2283.}
\bibitem{WEI}{  J. J. Weinstein and J. W. Negele,
                  Phys. Rev. Lett. {\bf 49}  (1982) 1016.}
\end{thebibliography}
\end{document}